\documentstyle[12pt]{article}
\newcommand{\tr}[1]{\,{\rm tr}\,#1\,}
\newcommand{\del}{\Delta }

\begin{document}
\title{
Elliptic Ruijsenaars-Schneider model via the Poisson reduction
of the Affine Heisenberg Double.
}
\author{G.E.Arutyunov,
\thanks{Steklov Mathematical Institute,
Vavilov 42, GSP-1, 117966, Moscow, Russia; arut@class.mi.ras.ru}\\
S.A.Frolov
\thanks{Steklov Mathematical Institute,
Vavilov 42, GSP-1, 117966, Moscow, Russia; frolov@class.mi.ras.ru}\\
and\\
P.B.Medvedev \thanks
{Institute of Theoretical and Experimental Physics,
 B.Cheremushkinskaja 25, 117259 Moscow, Russia}
}
\date {}
\maketitle
\begin{abstract}
It is shown that the elliptic Ruijsenaars-Schneider model can be obtained 
from the affine Heisenberg Double by means of the Poisson reduction 
procedure. The dynamical $r$-matrix naturally appears in the 
construction.  \end{abstract} \newpage

\section{Introduction}
The recent development \cite{Skl}-\cite{Sur1} in the theory of integrable 
many-body systems is mainly related with the discovery \cite{AT} of 
dynamical $r$-matrices, i.e. $r$-matrices depending on phase variables. 
One natural way to understand the origin of dynamical $r$-matrices is to 
consider the reduction procedure \cite{Arn}-\cite{OP2},\cite{ABT,AM}. 
In this approach one starts with an initial phase space ${\cal P}$ 
supplied with a symplectic action of some symmetry group. Considering 
a relatively simple invariant Hamiltonian and factorizing the 
corresponding dynamics by the symmetry group one gets a smaller phase 
space ${\cal P}_{red}$ with a nontrivial dynamics. Then the 
$L$-operator coming in the Lax representation $\frac{dL}{dt}=[M,L]$ 
appears as a specific coordinate on ${\cal P}_{red}$ while the 
dynamical $r$-matrix describes the Poisson (Dirac) bracket on the 
reduced space. 

At present the reduction procedure is elaborated for the majority of  
integrable many-body systems  and the corresponding $r$-matrices are 
derived. One of the most interesting exception is the 
elliptic Ruijsenaars-Schneider model \cite{R}. Recently two different 
dynamical $r$-matrices 
for this model were found in \cite{Sur1} and \cite{NKSR}. Both of these 
$r$-matrices were obtained by a direct calculation and the question of 
their equivalence still remains open. 

In this letter we apply the Poisson reduction procedure to the affine
Heisenberg Double (HD) \cite{Sem} and derive the elliptic 
Ruijsenaars-Schneider model with the dynamical $r$-matrix.
The reason to use the affine HD becomes apparent due to its
relation with integrable many-body systems of Calogero type.
As was shown in \cite{GN,GNH} the Calogero-Moser 
and the rational and trigonometric Ruijsenaars-Schneider  
hierarchies can be obtained by means of the reduction procedure from 
the cotangent bundle of an affine Lie group $T^* G(z)$ and from a 
finite dimensional Heisenberg Double. The affine Heisenberg Double 
may be regarded as a deformation of $T^* G(z)$ and therefore one can 
suggest that the affine HD is a natural candidate for the phase space 
standing behind the elliptic Ruijsenaars-Schneider system.  

The plan of the paper is as follows. In the second section we briefly  
describe the affine HD in terms of variables which are suitable for the 
reduction procedure. Then we fix the momentum map, corresponding to the 
natural action of the affine Poisson-Lie group on HD. The solution of 
the momentum map equation is shown to be equivalent to the 
$L$-operator of the elliptic Ruijsenaars-Schneider model. In the 
third section we study the Poisson structure of the reduced phase 
space and prove that it coincides with the one of the elliptic 
Ruijsenaars-Schneider model. 
The dynamical $r$-matrix naturally 
appears in our consideration and is equivalent to the one obtained in 
\cite{Sur1}. In the last section we show that the problem of solving 
the equations of motion is equivalent to the specific
factorization problem. 

In our presentation we omit the detailed description of the HD and 
the proof of some statements. The complete discussion will be given 
in a forthcoming publication.

\section {Affine Heisenberg Double}
The general construction of a Poisson manifold known as the 
Heisenberg double was elaborated in \cite{Sem}. We shall discuss the HD 
for the affine $\widehat{GL(N)}$. 
It is convenient to describe the Poisson structure of the 
affine HD in the following form. Let $A(x)$ and $C(x)$  
be formal Fourier series in a variable $x$ with values in $GL(N)$. The 
matrix elements of the harmonics of $A(x)$ and $C(x)$ can be regarded as 
generators of the algebra of functions on HD.  The Poisson 
structure on HD looks as follows:  
\begin{eqnarray} \frac1{\gamma}\{ 
A_1(x),A_2(y)\} &=&-r_{\mp}(x-y)A_1(x)A_2(y) 
-A_1(x)A_2(y)r_{\pm}(x-y)\nonumber\\ 
 &+& A_2(y)r_+(x-y-2\Delta)A_1(x)+ 
A_1(x)r_-(x-y+2\Delta)A_2(y),\nonumber\\
\frac1{\gamma}\{ C_1(x),C_2(y)\}
&=&-r_{\mp}(x-y)C_1(x)C_2(y)
-C_1(x)C_2(y)r_{\pm}(x-y)\nonumber \\
&+& C_2(y)r_+(x-y)C_1(x)+
C_1(x)r_-(x-y)C_2(y),\nonumber\\ 
\frac1{\gamma}\{ A_1(x),C_2(y)\}
&=&-r_-(x-y)A_1(x)C_2(y)
-A_1(x)C_2(y)r_-(x-y+2\Delta) \nonumber \\
&+& C_2(y)r_+(x-y)A_1(x)+
A_1(x)r_-(x-y+2\Delta)C_2(y), \nonumber \\
\frac1{\gamma}\{ C_1(x),A_2(y)\}
&=&-r_+(x-y)C_1(x)A_2(y)
-C_1(x)A_2(y)r_+(x-y-2\Delta) \nonumber \\
*&+& A_2(y)r_+(x-y-2\Delta)C_1(x)+
C_1(x)r_-(x-y)A_2(y),\nonumber 
\end{eqnarray} 
where $\gamma$ and $\Delta$ are complex numbers with $\mbox{Im}~\Delta>0$.
Here we use a standard tensor notation. The matrices $r_\pm (x)$ are 
defined by their Fourier series as
$$
r_+(x)=r_+ +P\sum_{n>0}e^{-inx},~~~
r_-(x)=r_- -P\sum_{n>0}e^{inx},
$$
where
$$
r_+ =\frac12 \sum_i E_{ii}\otimes E_{ii}+\sum_{i<j} E_{ij}
\otimes E_{ji},
$$
$r_- =-Pr_+P$ and $P$ is the permutation operator.
It can be easily checked that
$$
r_+ (x)-r_- (x)=2\pi P\delta (x)~~ \mbox{and}~~Pr_+ (x)P=-r_- (-x).
$$
In the region of convergence $r_\pm (x)$ coincide with the standard 
trigonometric $r$-matrix for the affine Lie algebra. 
The Poisson subalgebra generated by $A(x)$ was introduced in \cite{SR} 
to describe the Poisson structure of $\widehat{GL(N)}^*$.

Assuming the expansions
$$
A(x)=I+\gamma J(x)+\ldots,~~C(x)=g(x)+\ldots,~~
2\Delta=\gamma k,
$$
where $k$ is a (fixed) central charge,  in the 
deformation limit $\gamma\to 0$ we recover the 
standard Poisson structure on the cotangent 
bundle $T^*\widehat{GL(N)}$ over the level k centrally extended current 
group $\widehat{GL(N)}$. 

The action of the current group $\widetilde{GL(N)}$ on HD:
\begin{eqnarray*} 
A(x)&\to &T^{-1}(x-\del )A(x)T(x+\del ),\\
C(x)&\to &T^{-1}(x-\del )C(x)T(x-\del )
\end{eqnarray*} 
is Poissonian. Thereby, we can consider the Poisson reduction 
of HD over the action of $\widetilde{GL(N)}$.

The momentum map taking value in 
$\widetilde{GL(N)}^*$ reads as follows:
$$
M(x)=A^{-1}(x-\del )C(x-\del )A(x-\del )C^{-1}(x+\del ).
$$
It is easy to check that $M(x)$ does generate the action of the current 
group.
We fix the value of $M(x)$ as:
\begin{equation} 
M(x)=e^{ih}( 1-2\pi i \delta_\varepsilon (x)
\frac{1-e^{-ix}}{i} K).
\label{m}\end{equation} 
Here $h$ and $\varepsilon$ are arbitrary complex numbers, 
$$
\delta_\varepsilon (x)=\frac 1 \varepsilon \left(
\theta (x+\frac\varepsilon 2) -\theta (x-\frac\varepsilon 2) \right)
=\frac1{2\pi i \varepsilon }\sum^{n=+\infty}_{n=-\infty} \frac1n 
(e^{in\frac\varepsilon 2} -e^{-in\frac\varepsilon 2}) e^{inx}, 
$$ 
and 
$K$ is a constant matrix $K=e\otimes e^t$, where $e$ is the 
$N$-dimensional vector with entries $e_i = 1/\sqrt{N}$
\footnote{It is worthwhile to note that in the deformation limit 
$\gamma \to 0, ~~\frac h\gamma \to {\mbox const}, ~~\frac \varepsilon\gamma 
\to {\mbox const}$ the constraint (\ref{m}) reduces to the one used in
\cite{GN} to get the trigonometric Ruijsenaars model.}

Although eq.(\ref{m}) can be solved for any value of $\varepsilon$,
the reduced phase space remains to be 
infinite dimensional after performing the factorization procedure .  
To extract a finite dimensional phase space let us carry out the 
following trick. By multiplying the both sides of (\ref{m}) on
$ C(x+\del)$, one gets
\begin{eqnarray}
C(x+\del)-e^{-ih}A^{-1}(x-\del)C(x-\del)A(x-\del)
=2\pi i \delta_\varepsilon (x) K\frac{1-e^{-ix}}{i}C(x+\del).
\label{fe}
\end{eqnarray}
The l.h.s. of this equation
does not have any explicit dependence on $\varepsilon$.
As to the r.h.s., when $\varepsilon$ tends to zero, 
$\delta_\varepsilon (x)$ tends to $\delta(x)$ and the 
r.h.s. is well defined only if the function 
$\frac{1-e^{-ix}}{i}C(x+\del)$ is well 
defined at $x=0$. Hence, this equation 
can be solved only for meromorphic functions $C(x+\del)$  
 with poles of the first order. In this case 
$\lim_{\varepsilon \to 0} 
\delta_\varepsilon (x)\frac{1-e^{-ix}}{i}C(x+\del)= 
\delta (x){\mbox {Res}}_{x=0} C(x+\del)$.

So we define the constraint surface as being the solution of 
the equation
\begin{eqnarray}
C(x+\del)- e^{-ih}A^{-1}(x-\del)C(x-\del)A(x-\del)
=2\pi i\delta (x)K{\mbox {Res}}_{x=0} C(x+\del)
\nonumber
\end{eqnarray}
and in the following we shall explore solutions of this equation.

We start with the following difference equation
\begin{equation}
C(x+\Delta )-
e^{-ih}D^{-1} C(x-\Delta ) D=2\pi i \delta(x)Y,
\label{e}
\end{equation} 
where $D$ is a constant diagonal matrix and $Y$ is an arbitrary
constant matrix.
Performing the Fourier expansion we get a solution of (\ref{e}) in the form
$$
C(x)=i\sum_{ij}\sum_{n=-\infty}^{n=\infty}
\frac{e^{inx}}{e^{in\del}-e^{-is_{ij}}e^{-in\del}}Y_{ij}E_{ij},
$$
where we use the notation $s_{ij}=h+q_{ij}$, 
$D=e^{iq}$, $q_{ij}=q_i -q_j$. 

\noindent It is useful to introduce the function of
two complex variables
$$
w(x,s)=i\sum_{n=-\infty}^{n=\infty}\frac{e^{inx}}{e^{in\del 
 }-e^{-is}e^{-in\del }}.  $$ It is clear that $w(x,s)$ is a 
meromorphic function of $s$ for any $x:  |\mbox{Im}~ x|<\mbox{ Im}~ 
\del$ and has two obvious properties \begin{enumerate} \item 
$w(x,s+2\pi)=w(x,s)$, \item 
$w(x,s+2\Delta )=e^{i\Delta -ix}w(x,s)$.  \end{enumerate} 
Moreover, as a function of $s$ it has simple poles at $0,\pm 2\pi, 
\pm 4\pi, \ldots$ and $\pm 2\Delta, \pm 4\Delta, \ldots$, ${\mbox 
{Res}}_{s=0} w=1$.  By these data $w$ is uniquely defined as: 
\begin{equation} 
w(x,s)= 
\frac{\sigma(s+x-\Delta )}
{\sigma(x-\Delta )\sigma(s)}
e^{-\frac{\zeta(\pi)}{\pi}(x-\Delta )s}.
\nonumber
\end{equation} 
Here $\sigma (x)$ and $\zeta (x)$ are the Weierstrass $\sigma$- 
and $\zeta$-functions with periods equal to $2\pi$ and $2\Delta$.
Thus, equation (\ref{e}) has the unique solution 
\begin{equation}
C(x)=
\sum_{ij}
\frac{\sigma(q_{ij}+h+x-\del )}
{\sigma(x-\del )\sigma(q_{ij}+h)}
e^{-\frac{\zeta(\pi)}{\pi}(x-\del )(h+q_{ij})}Y_{ij}E_{ij} =
\sum_{ij} w(x,s_{ij}) Y_{ij}E_{ij}.
\label{fin1}
\end{equation} 

Now we turn to the momentum map equation 
\begin{equation}
C(x+\del )-
e^{-ih}A^{-1}(x-\del )C(x-\del )A(x-\del )
=2\pi i KZ\delta(x),
\label{fg}
\end{equation} 
where $Z={\mbox {Res}}_{x=0}C(x+\Delta )$. 

By using a generic gauge 
transformation we can diagonalize the field $A$. Then equation (\ref{fg}) 
takes the form of eq.(\ref{e})
\begin{equation} 
C'(x+\del)- 
e^{-ih}D^{-1}C'(x-\del)D
= 2\pi i K'Z'\delta(x),
\nonumber
\end{equation} 
where 
$$
A(x)=T(x-\del)DT^{-1}(x+\del),~~
C(x)=T(x-\del)C'(x)T^{-1}(x-\del)
$$
for some $T$
and $Z'={\mbox {Res}}_{x=\del}C'(x)$. We also have
$$
K'=T^{-1}(0)KT(0)=T^{-1}(0)e\otimes e^tT(0)=f\otimes v^t,~~~<f,v>=1
$$
i.e. $f=T^{-1}(0)e$ and $e^tT(0)=v^t$.
According to (\ref{fin1}) we find
$$
C'(x)=\sum_{ij}
w(x,s_{ij})(K'Z')_{ij}E_{ij}.
$$
Taking the residue of $C'(x)$ at the point $x=\Delta $ we arrive
at the compatibility condition
$$
Z'=K'Z'=f\otimes v^t Z',~~~<f,v>=1. 
$$
The solution of this equation is $Z'=f\otimes g^t$, where $g$ is an 
arbitrary vector.
Now it is easy to find $Z$:
$$Z=T(0)Z'T^{-1}(0)=T(0)f\otimes g^tT^{-1}(0)=
e\otimes g^tT^{-1}(0)
\equiv e\otimes b^t
$$
Thus, we get
\begin{equation} 
C(x+\Delta )-
e^{-ih}A^{-1}(x-\Delta )C(x-\Delta )A(x-\Delta )
= 2\pi i (e\otimes e^t) (e\otimes b^t)  \delta(x),
\label{fg11}
\end{equation} 
where $ e\otimes b^t$ is a residue of $C(x)$ at $x=\Delta$. 

To summarize, eq.(\ref{fg}) has a solution for any field $A$ 
and for any field $C$, having a residue at $x=\del$ of the form
$ e\otimes b^t$. For a fixed field $A$ and a vector $b$ this solution 
is unique.
Note that, in general,
$<b,e>\neq 1$. The form of the r.h.s. of (\ref{fg11}) shows that
the isotropy group of this equation is
$$
G_{isot}=\{T(x)\subset G(x)~~|~~ T(0)e=\lambda e, 
\lambda\in {\bf C}\}.
$$ 
This group transforms a solution of 
(\ref{fg11}) into another one, so the reduced phase space is defined 
as 
$$
{\cal P}_{red}=\frac{\mbox{all solutions of (\ref{fg}})}{G_{isot}}.
$$
Since the group $G_{isot}$ is large enough to diagonalize the field
$A$, we can parametrize the reduced phase space by the section
$(D,L)$, where $L$ is a solution of (\ref{fg}) with $A=D$. 
One can easily see that ${\cal P}_{red}$ is finite dimensional
and it's dimension is 
equal to $2N$, i.e. $N$ coordinates of $D$ plus $N$ coordinates of the 
vector $b$.
Due to eq.(\ref{fin1}) the corresponding $L$-operator has the following form:
\begin{equation}
L(x)=\sum_{ij}
\frac{\sigma(q_{ij}+h+x-\del )}
{\sigma(x-\del )\sigma(q_{ij}+h)}
e^{-\frac{\zeta(\pi)}{\pi}(x-\del )(h+q_{ij})} 
e_{i}b_j E_{ij}.
\label{l}\end{equation}
Multiplying $L(x)$ by the function $
\frac {\sigma(x-\del )\sigma(h)}{\sigma(x-\del +h)} 
e^{\frac{\zeta (\pi )}{\pi} (x-\del)h}$,
 performing the gauge transformation by means of the diagonal matrix $
e^{\frac{\zeta (\pi )}{\pi} (x-\del)q} 
$, and making the shift $x\to x+\del$
we obtain
the $L$-operator of the elliptic Ruijsenaars-Schneider model:
\begin{equation}
L^{{\em Ruij}}(x)=
\frac {\sigma(x )\sigma(h)}{\sigma(x+h)} 
e^{\frac{\zeta (\pi )}{\pi} xh}
e^{\frac{\zeta (\pi )}{\pi} xq}
L(x+\del )
e^{-\frac{\zeta (\pi )}{\pi} xq}
\label{rg}
\end{equation}

Let us briefly discuss the Hamiltonian. It is well known that the simplest 
nontrivial Hamiltonian invariant with respect to the 
action of the current group is given by:
\begin{equation}
H=\int dx \tr C(x),
\label{HH}
\end{equation}
where $\alpha$ is a constant. 
It is not difficult to show that on the reduced phase space 
\begin{equation}
H_{red}=\frac{2\pi i}{\sqrt{N}(1-e^{-ih})} \sum_{i=1}^N b_i 
\label{h}
\end{equation}
that is up to a constant nothing but the simplest Hamiltonian of the 
elliptic Ruij\-se\-naars-Schneider model. 
\section{The Poisson structure on the reduced space}
In this section we are going to prove that the Poisson structure on the 
reduced phase space does coincide with the Poisson structure of the
elliptic Ruijsenaars-Schneider model. In other words we need Poisson 
brackets for the coordinates $D$-s and $b$-s. According to the general 
Dirac construction one should find a gauge invariant extension 
(we mean the invariance under the action of $G_{isot}$) of 
functions on the reduced phase space ${\cal P}_{red}$ to a vicinity of 
${\cal P}_{red}$ and then calculate the Dirac bracket. 

One can easily write down the gauge invariant extension for the matrices
$D$ and $L(x)$ while the bracket for the coordinates $D_i$ and $b_i$ can 
be extracted from the bracket for $D$ and $L(x)$. 
This extension looks as follows:
\begin{equation}
D\rightarrow 
D[A]=T^{-1}[A](x-\Delta )A(x)T[A](x+\Delta )
\label{fac}
\end{equation} 
\begin{equation}
L(x) \rightarrow {\cal L}[A,C](x)
=T^{-1}[A](x-\Delta )C(x)T[A](x-\Delta ).
\nonumber
\end{equation} 
Some comments are in order. Eq. (\ref{fac}) is a solution of 
the factorization problem for $A(x)$. Generally this solution is not unique
but we fix the matrix $T[A]$ by the boundary condition 
$T[A](0)e= e   $ that kills 
the ambiguity and makes (\ref{fac}) to be correctly defined.
It is obvious that on ${\cal P}_{red}$: $T[A]=1$ and ${\cal 
L}[A,C](x)=L(x)$. 

We start with the calculation of the Poisson bracket for ${\cal L}(x)$ and 
${\cal L}(y)$. We omit the discussion of the contribution from the second 
class constraints to the Dirac bracket till the end of the section.  By 
definition, one has 
\begin{eqnarray} 
\{{\cal L}_1,{\cal L}_2\}_{{\cal 
P}_{red}}&=& \left(\{T_1,T_2\}L_1L_2 -L_2 \{T_1,T_2\}L_1-L_1 \{T_1,T_2\}L_2 
\right. \nonumber \\
&+&
L_1L_2 \{T_1,T_2\}
+\{C_1,C_2\}-\{T_1,C_2\}L_1-\{C_1,T_2\}L_2
\nonumber \\
&+&L_2\{C_1,T_2\}+L_1\{T_1,C_2\}
\left. \right) |_{{\cal P}_{red}}
\label{gh} 
\end{eqnarray}
Here we took into account that $T[A]|_{{\cal P}_{red}}=1$.

Let us first calculate
$$
\{C_{ij}(x),T_{kl}(y)\}=\sum_{m,n}
\int dz \{C_{ij}(x),A_{mn}(z)\} \frac{\delta T_{kl}(y)}
{\delta A_{mn}(z)}.  
$$ 
Performing the variation of 
the both sides of (\ref{fac}), we get
\begin{equation}
X(x)=t(x-\Delta )D-Dt(x+\Delta )+d,
\label{sd}
\end{equation}
where $X(x)=\delta A(x)$, $t(x)=\delta T(x)$
and $d=\delta D$.

\noindent The general solution of (\ref{sd}) is
\begin{equation}
t(x)=Q-\frac{1}{2\pi i}\sum_{i, j}
\int dz \frac{1}{D_i}w(x-z,q_{ij})X_{ij}(z) E_{ij}.
\nonumber
\end{equation}
Here $Q$ is some constant diagonal matrix and the function
$w(x,0)$ should be understood as 
$$ 
w(x,0)=\lim_{\varepsilon \to 
0}(w(x,\varepsilon) - \frac{i}{1-e^{-i\varepsilon}}) 
=\zeta(x-\del)-\frac{\zeta(\pi)}{\pi}(x-\del)-\frac{i}{2}.
$$
Note that these functions solve the equations
$$
\frac{1}{2\pi i}(w(x+\Delta ,q_{ij})-
e^{-iq_{ij}}w(x-\Delta ,q_{ij}))=\delta(x )-\frac{1}{2\pi}
\delta_{ij}.
$$
The solution $t(x)$ obeying the condition $t(0)e=0$ has the 
following form
\begin{equation}
t(x)=\frac1{2\pi i}\sum_{i,j}\int dz ( \frac1{D_i}w(-z,q_{ij})
X_{ij}(z)E_{ii}-
\frac1{D_i}w(x-z,q_{ij})
X_{ij}(z)E_{ij} )
\label{t}\end{equation}
Performing the variation of eq.(\ref{t}) with respect to $X_{mn}(z)$ one gets
$$
\frac{\delta T_{kl}(x)}{\delta A_{mn}(z)}|_{{\cal P}_{red}}
\equiv Q^{kl}_{mn}(x+\del ,z)
=\frac1{2\pi i}  \frac1{D_k}\left( w(-z,q_{kn})
\delta_{kl}\delta_{km} -w(x-z,q_{kl})
\delta_{km}\delta_{ln}\right)
$$
Thus on the reduced space we obtain
$$
\frac 1\gamma \{C_1(x),T_2(y-\del )\}|_{red}=
\kappa_{12}(x,y)L_1(x)-L_1(x)\omega_{12}(x,y),
$$
where
\begin{eqnarray*}
\kappa_{12}(x,y)=
{\mbox tr}_3\int dz
(D_3r_{+}^{13}(x-z-2\Delta)-
r_{+}^{13}(x-z)D_3)Q_{23}(y,z),\\
\omega_{12}(x,y)=
{\mbox tr}_3\int dz
(D_3r_{+}^{13}(x-z-2\Delta)-
r_{-}^{13}(x-z)D_3)Q_{23}(y,z).
\end{eqnarray*}
We also get
$$
\frac 1\gamma \{T_1(x-\del),C_2(y)\}=-P\kappa_{12}(y,x)P L_2(y)+
L_2(y)P\omega_{12}(y,x)P.
$$
By using the relation
$$
D_j Q^{kl}_{ij}(x,z)-D_i Q^{kl}_{ij}(x,z-2\Delta)=
\delta(x-z)\delta_{ik}\delta_{jl}-\delta(z-\Delta )
\delta_{ik}\delta_{kl}\equiv  S^{kl}_{ij},
$$
we find
\begin{eqnarray*} 
\kappa _{ij~kl}(x,y)&=&-r_+ (x-y)_{ij~kl} +
\sum_m r_+ (x-\Delta  )_{ij~km} \delta_{kl} \\
\omega _{ij~kl}(x,y)&=&k_{ij~kl}(x,y)+2\pi D_i Q_{ji}^{kl}(y,x).
\end{eqnarray*} 

Recall that
\begin{eqnarray*}
\frac1{\gamma}\{ C_1(x),C_2(y)\}
&=&-r_{\mp}(x-y)C_1(x)C_2(y)
-C_1(x)C_2(y)r_{\pm}(x-y)\\
&+& C_2(y)r_+(x-y)C_1(x)+
C_1(x)r_-(x-y)C_2(y), 
\end{eqnarray*} 
Substituting $\{ C,T\}$, $\{ T,C\}$ and $\{ C,C\}$ brackets into 
(\ref{gh}) we can rewrite the $\{ {\cal L},{\cal L}\}$ bracket in the 
following form:  
\begin{eqnarray} 
\frac 1\gamma \{{\cal L}_1(x),{\cal L}_2(y)\}|_{red} 
&=&-L_1(x)L_2(y)k^+(x,y)-k^-(x,y)L_1(x)L_2(y)\nonumber\\ &+&
L_1(x)s^-(x,y)L_2(y)+L_2(y)s^+(x,y)L_1(x),
\label{nn}
\end{eqnarray} 
where
\begin{eqnarray}
k^-(x,y)&=&r_-(x-y)+
\kappa_{12}(x,y)-P\kappa_{12}(y,x)P-
\{T_1(x-\del),T_2(y-\del)\},\nonumber \\
k^+(x,y)&=& r_+(x-y)+
\omega_{12}(x,y)-P\omega_{12}(y,x)P-
\{T_1(x-\del),T_2(y-\del)\},\nonumber \\
s^-(x,y)&=&r_-(x-y)
+\omega_{12}(x,y)-P\kappa_{12}(y,x)P-
\{T_1(x-\del),T_2(y-\del)\},\nonumber \\
s^+(x,y)&=&r_+(x-y)+
\kappa_{12}(x,y)-P\omega_{12}(y,x)P-
\{T_1(x-\del),T_2(y-\del)\}.\nonumber 
\end{eqnarray}
It is easy to find $Pk^{\pm}(x,y)P=
- P \delta(x-y)-k^{\pm}(y,x)$ and
$Ps^{\pm}(x,y)P=\pm s^{\mp}(y,x)$.
We also have one more important identity
$$
k^{+}(x,y)+k^{-}(x,y)=
s^{+}(x,y)+s^{-}(x,y).
$$

To complete the calculation we should find the bracket 
$\{T_{ij}(x-\Delta  ) , T_{kl}(y-\Delta  )\}$
on the reduced space.
The straightforward manipulations lead to a divergent result.
By this reason we define this bracket as follows:
$$
\{T_{ij}(x-\Delta  ) , T_{kl}(y-\Delta  )\}
=\frac 1 2 \lim_{\varepsilon \to 0} \left(
\{T_{ij}(x-\Delta  ) , T^{\varepsilon}_{kl}(y-\Delta  )\}
+\{T^{\varepsilon}_{ij}(x-\Delta  ) , T_{kl}(y-\Delta  
)\}\right) 
$$ 
where $T^{\varepsilon}_{kl}(x)$ is defined as a solution of 
the factorization problem with the boundary condition $T(\varepsilon 
)e=e$. We have 
\begin{eqnarray*} 
\{T_{ij}(x-\Delta ) , 
T_{kl}^\varepsilon(y-\Delta  )\}= \int dz dz' 
Q^{ij}_{mn}(x,z)Q^{kl~\varepsilon}_{sp}(y,z')\{A_{mn}(z),A_{sp}(z')\}
\end{eqnarray*} 
$$
=\gamma \int dz dz'\left(
-r_{+}(z-z')_{mn~sp}(D_n Q^{ij}_{mn}(x,z)-D_m Q_{mn}^{ij}(x,z-2\Delta))
S^{kl~\varepsilon}_{sp}(y,z')\right.
$$
$$
\left. -2\pi P_{mn~sp}\delta(z-z'+2\Delta)D_m Q^{ij}_{mn}(x,z)
S^{kl~\varepsilon}_{sp}(y,z')\right)
$$

One can prove the cancellation of the singularities as $\varepsilon\to 0$. 
The result for the bracket $\{T,T\}$ is 
\begin{eqnarray*} 
&& \frac 1\gamma 
\{T_{ij}(x-\del ) , T_{kl}(y-\del )\} \\ 
&&= -r_+ (x-y)_{ij~ kl} + \sum_{m} r_+ (x-\del )_{ij~ km}\delta_{kl}
+\sum_{m} r_- (\del -y)_{im~ kl}\delta_{ij}\\
&&+ \frac 1 i w(x-y+\del ,q_{ik})\delta_{jk}\delta_{il}
-\frac 1 i w(x,q_{ik})\delta_{jk}\delta_{kl}
+\frac 1 i w(y,q_{ki})\delta_{ij}\delta_{il} \\
&&+\frac 1 2 \delta_{ij}\delta_{ik}\delta_{il} +\frac 1i(\zeta 
(q_{ik}) - \frac{\zeta(\pi)}{\pi} q_{ik}  
)\delta_{ij}\delta_{kl}(1-\delta_{ik})-
\frac 1 2 \sum_{a<b}(E_{ab}-E_{ba})_{ik}\delta_{ij}\delta_{kl}
\end{eqnarray*} 
Combining all the pieces together and taking into account the identity
$ e^{-is} w(x,s)=-w(-x,-s)$ 
we get the following expression for the coefficients:
\begin{eqnarray*} 
k^- _{ij~kl}(x,y )&=&
-\frac 1 i \zeta (q_{ik}) \delta_{ij}\delta_{kl} (1-\delta_{ik})
-\frac 1 i \left( \zeta (x-y) +\zeta (y-\del )-\zeta (x-\del )\right)
\delta_{ij}  \delta_{ik} \delta_{il}
\\
&-&\frac 1 i \left( w( x-y + \del ,q_{ik}) 
\delta_{il}\delta_{jk} + w(y,q_{ki})\delta_{il}\delta_{ij}
-w(x,q_{ik})\delta_{jk} \delta_{kl}  \right) (1-\delta_{ik})\\
&+& 
 \frac 1 i \frac{\zeta (\pi)}{\pi} 
q_{ik}  \delta_{ij}\delta_{kl}
+\frac 1 2 \sum_{a<b}(E_{ab}-E_{ba})_{ik} \delta_{ij}\delta_{kl}
 \\
k^+ _{ij~kl}(x,y )&=&
\frac 1 i \left( \zeta (x-y) -\frac{\zeta(\pi)}{\pi}(x-y)   
\right) \delta_{ij}  \delta_{ik} \delta_{il}
\\
&+&\frac 1 i\left( 
w(x-y +\del ,q_{ik})\delta_{jk}\delta_{il} 
-\zeta(q_{ik})  \delta_{ij}\delta_{kl}\right) 
(1-\delta_{ik}) \\
&+&\frac 1 i \frac{\zeta(\pi)}{\pi}q_{ik} 
\delta_{ij}\delta_{kl} 
+ \frac 1 2 \sum_{a<b}(E_{ab}-E_{ba})_{ik} 
\delta_{ij}\delta_{kl}  \\
s^-_{ij~kl}(x,y)&=&
-\frac 1 i \left( \zeta (y-\del ) -\frac{\zeta(\pi)}{\pi}(y-\del )   
\right) \delta_{ij}\delta_{ik} \delta_{il} 
\\
&-&\frac 1 i\left( 
w(y ,q_{ki})\delta_{ij}\delta_{il} 
+\zeta(q_{ik})  \delta_{ij}\delta_{kl}\right) 
(1-\delta_{ik}) \\
&+&\frac 1 i \frac{\zeta(\pi)}{\pi}q_{ik} 
\delta_{ij}\delta_{kl} 
+ \frac 1 2 \sum_{a<b}(E_{ab}-E_{ba})_{ik} 
\delta_{ij}\delta_{kl}  \\
s^+_{ij~kl}(x,y)&=&
\frac 1 i \left( \zeta (x-\del ) -\frac{\zeta(\pi)}{\pi}(x-\del )   
\right) \delta_{ij}\delta_{ik} \delta_{il} 
\\
&+&\frac 1 i\left( 
w(x ,q_{ik})\delta_{jk}\delta_{kl} 
-\zeta(q_{ik})  \delta_{ij}\delta_{kl}\right) 
(1-\delta_{ik}) \\
&+&\frac 1 i \frac{\zeta(\pi)}{\pi}q_{ik}\delta_{ij}\delta_{kl} 
+ \frac 1 2 \sum_{a<b}(E_{ab}-E_{ba})_{ik} 
\delta_{ij}\delta_{kl}  \\
\end{eqnarray*}
It is instructive to note that one can check by direct calculation
that the term 
$\frac{\zeta(\pi)}{i\pi}q_{ik}\delta_{ij}\delta_{kl}$
in the expressions obtained for $k$-s and $s$-s 
does not contribute to the bracket $\{ {\cal L},{\cal L}\}$.

Remind that (see eq.(\ref{l})):
\begin{equation}
L_{ii} (x) = \frac 1{\sqrt{N}} w(x,h)b_i ,
\nonumber 
\end{equation}
so to obtain the bracket $\{ b_i ,b_j \}$ it is sufficient to examine the
$\{ L_{ii}, L_{jj}\}$ bracket only. The crucial point which can be checked by 
the direct calculation is that the bracket of ${\cal L}_{ii}$ with the 
constraint (\ref{fe}) vanishes on ${\cal P}_{red}$ in the limit 
$\varepsilon \to 0$. Thus, there is no contribution from the Dirac term to 
the $\{ L_{ii}, L_{jj}\}$ bracket. 

By substituting the expressions obtained above for $k$ and $s$ into 
eq.(\ref{nn}), one gets for $i\ne j$
\begin{equation}
\frac 1\gamma \{ L_{ii}(x), L_{jj}(y)\}=
\frac 1 i L_{ji}(x)L_{ij}(y)w(x-y+\del ,q_{ij}) 
- \frac 1 i L_{ij}(x)L_{ji}(y)w(x-y+\del ,q_{ji}). 
\nonumber
\end{equation}
It follows from this equation that
\begin{equation}
\frac i\gamma \{ b_i , b_j \}= b_i b_j 
\frac{w(x,s_{ji})w(y,s_{ij})w(x-y+\del,q_{ij})
-w(x,s_{ij})w(y,s_{ji})w(x-y+\del,q_{ji})}{w(x,h)w(y,h)}.
\label{bb} 
\end{equation}
By using one of the known elliptic identities \cite{NKSR}\footnote
{It is interesting to note that this identity can be easily obtained from 
the $x,y$-independence condition for the r.h.s. of eq.(\ref{bb}). }, 
we get 
\begin{equation} 
\frac i\gamma \{ b_i , b_j \}= b_i b_j 
(2\zeta (q_{ij})-\zeta (q_{ij}+h) -\zeta (q_{ij}-h)).  
\nonumber 
\end{equation}

To complete the examination of the Poisson structure on the reduced 
phase space one should find the bracket $\{ {\cal L}, D\}$ and
$\{ D, D\}$. 
Performing the straightforward but rather tedious calculations 
following the same line as above, we find
\begin{equation}
\{D[A]_1,D[A]_2\}|_{red}=0,
\nonumber
\end{equation}
\begin{equation}
\frac 1\gamma \{{\cal L}(x)_1,D[A]_2\}|_{red}=- 
\sum_{i,j}L_{ij}(x)D_j E_{ij}\otimes E_{jj}. 
\label{LD} 
\end{equation}
It is worthwhile to point out that there are no Dirac terms in these 
brackets because $D[A]$ is invariant with respect to the action of the 
whole affine group $\widetilde{GL(N)}$.

Now for the reader's convenience we list the Poisson brackets obtained in 
terms of the coordinates on ${\cal P}_{red}$
\begin{eqnarray}
\{q_i ,q_j \}&=&0 \nonumber \\
\frac i \gamma \{q_i ,b_j \}&=&b_j \delta_{ij} \nonumber \\
\frac i\gamma \{ b_i , b_j \}&=& b_i b_j 
(2\zeta (q_{ij})-\zeta (q_{ij}+h) -\zeta (q_{ij}-h)).  
\label{pb}
\end{eqnarray}
One can see that the dynamical system defined by (\ref{pb}) and (\ref{h})
is nothing but the elliptic Ruijsenaars-Schneider model.

In \cite{Sur1} the dynamical $r$-matrix for $L^{ Ruij}$ was obtained by 
direct calculation with the help of the Poisson structure (\ref{pb}).
Comparing the $r$-matrix coefficients $k$ and $s$ with the ones in 
\cite{Sur1} we see that, in fact, they differ by the tensor $
\frac 1 2 \sum_{a<b}(E_{ab}-E_{ba})_{ik} 
\delta_{ij}\delta_{kl} $. However, in our calculations of the bracket 
$\{{\cal L},{\cal L}\}$ we ignored the contribution from the Dirac term.
We conjecture that just the Dirac term is responsible for cancelling this 
tensor. 

\section{Equations of motion}
The equations of motion for the Hamiltonian (\ref{h}) are given by
\begin{equation}
\dot{D}=\{\mbox{tr}L(x),D\}=-\gamma L(x)_{diag}~D
\label{MD}
\end{equation} 
and
\begin{equation}
\dot{L}(y)=\{\mbox{tr}L(x),L(y)\}=[L(y),M(x,y)],
\label{ML}
\end{equation} 
where 
\begin{equation}
M(x,y)=-\gamma i \sum_{kl}\left(
w(x,-q_{kl})L(x)_{kl}E_{kk}-
w(x-y+\Delta ,-q_{kl})L(x)_{kl}E_{kl}\right).
\label{Mst}
\end{equation} 
Here we use eq.(\ref{nn}) and the explicit form of $k$ and $s$. Since
$\mbox{tr}L(x)$ is invariant function the contribution from the Dirac term
vanishes. For the reader's convenience we note that
by using the elliptic function identities \cite{NKSR} one can 
rewrite $M\equiv M(x+\Delta ,y+\Delta )$ in the following 
form 
\begin{eqnarray}
\label{ME}
M&=&\frac{\gamma}{i}l(x,h)\left(                
\frac{\zeta(x+h)-\zeta(x-y)}{l(y,h)}L(y+\Delta )
-(\zeta(x+h)-\zeta(x))(\sum_i b_i)I \right. \\
&+& 
\sum_k E_{kk}\sum_{i\neq k}(\zeta(q_{ik})-\zeta(q_{ik}-h))b_i
-\frac{\zeta(\pi)}{\pi}\sum_k b_k E_{kk}\\
&+& \left.
\sum_{k\neq l}\frac{\zeta(q_{kl})-\zeta(q_{kl}+y+h)}{l(y,h)}
L_{kl}(y+\Delta )E_{kl}\right),
\end{eqnarray}
where we have introduced $l(x,h)=w(x+\Delta ,h)$.
The first two terms in (\ref{ME}) are irrelavant, so $M$ coincides with
the standard $M$-matrix of the elliptic RS system.

We show that the general
solution of the equations of motion for the elliptic Ruijsenaars-Shneider 
model is given by
\begin{equation}
D(t)=D[e^{-2\pi \gamma L_0(x)t}D_0],
\label{m1}
\end{equation} 
where $D\equiv D[A]$ denotes the solution of the factoriztion problem 
(\ref{fac}):
\begin{equation}
A(x)=T(x-\Delta )D[A]T(x+\Delta )^{-1}
\label{f1}
\end{equation} 
and $D_0$, and $L_0(x)$ are the coordinates and the 
$L$-operator at $t=0$ respectively. 

To prove (\ref{m1}) we start with calculating the derivative $\dot{D}(t)$:
\begin{equation}
\dot{D}(t)=\int dz 
\frac {\delta D[A]}{\delta A_{ij}(z)}|_{A=A_t}\frac{d(A_t)_{ij}(z)}{dt},
\label{s}
\end{equation} 
where $A_t(x)=e^{-2\pi \gamma L_0(x)t}D_0$. One can find the derivative
$\frac {\delta D[A]}{\delta A_{ij}(z)}$ by performing the variation
of eq.(\ref{f1}):
\begin{equation}
(T^{-1}(x-\Delta)\delta T(x-\Delta))
D[A]-D[A](T^{-1}(x+\Delta)\delta T(x+\Delta))+\delta D=X(x),
\label{we}
\end{equation} 
where the notation 
$X(x)=T^{-1}(x-\Delta )\delta A(x)T(x+\Delta )$
was introduced. In contrast to (\ref{sd}) in eq.(\ref{we}) we do not
impose the constraint $T=1$.

Now we solve (\ref{we}) for $\delta D$:
\begin{equation}
\delta D=\int \frac{dx}{2\pi} X(x)_{kk}E_{kk}.
\label{sd1}
\end{equation} 
Eq.(\ref{we}) also allows one to find the matrix
\begin{equation}
T^{-1}(x)\delta T(x)=\sum_{k,l}
\int\frac{dz}{2\pi i}
\left(\frac{1}{D_k} w(-z,q_{kl})X(z)_{kl}E_{kk}-
\frac{1}{D_k}w(x-z,q_{kl})X(z)_{kl}E_{kl}\right)
\label{sd2}
\end{equation} 
that will be used in the sequel. From (\ref{sd1}), (\ref{sd2}) we find
\begin{equation}
\frac {\delta D[A]_{kk}}{\delta A_{ij}(z)}=\frac{1}{2\pi}
T^{-1}_{ki}(z-\Delta )T_{jk}(z+\Delta )
\label{fgg}
\end{equation} 
and
\begin{eqnarray}
\label{fg1}
\left(T^{-1}(x)\frac{\delta T(x)}{\delta A_{ij}(z)}\right)_{kl}&=&
\frac{\delta_{kl}}{2\pi i}\sum_s \frac{w(-z,q_{ks})}{D_k}
T^{-1}_{ki}(z-\Delta )T_{js}(z+\Delta )\\
\nonumber
&-& \frac{1}{2\pi i}\frac{w(x-z,q_{kl})}{D_k}
T^{-1}_{ki}(z-\Delta )T_{jl}(z+\Delta ).
\end{eqnarray} 
Substituting (\ref{fg1}) in eq.(\ref{s}) and taking into account
$\dot{A_t}(x)=-2\pi \gamma L_0(x)A_t(x)$, we get
\begin{equation}
\dot{D}(t)_{kk}=-\gamma \int dz 
T^{-1}_{ki}(z-\Delta )L_0(z)_{im}T_{mn}(z-\Delta )
T^{-1}_{ns}(z-\Delta )A_t(z)_{sj}T_{jk}(z+\Delta )
\label{q1}
\end{equation} 
that with the help of (\ref{f1}) reads as follows
\begin{equation}
\dot{D}(t)=-\gamma \int dz 
(T^{-1}(z-\Delta )L_0(z)T(z-\Delta ))_{\mbox{diag}}~D(t).
\label{q2}
\end{equation} 
The last formula implies the notation
\begin{equation}
\hat{L}_t(x)=T^{-1}(x-\Delta )(t)L_0(x)T(x-\Delta )(t)
\label{q3}
\end{equation} 
that provides the Lax representation 
$\frac{d}{dt}{\hat{L}}_t(x)=[\hat{L}_t(x),\hat{M}(x)]$
with $\hat{M}(x)=T^{-1}(x-\Delta )\dot{T}(x-\Delta )$.

Let us show that the Lax operator $\hat{L}_t(x)$ coincides with the 
$L$-operator of the elliptic Ruijsensaars-Shneider model. To this end
we calculate explicitly $\hat{M}(x)$. We have
$$
\hat{M}_{kl}(x)=\int 
\left(T^{-1}(x-\Delta )
\frac{\delta T(x-\Delta )}{\delta A_{ij}(z)}\right)_{kl}|_{A=A_t}
\frac{d A_{t}(z)_{ij}}{dt}
$$
Substituting (\ref{fg1}) and using the relation $e^{-is}w(x,s)=-w(-x,-s)$
we get 
\begin{equation}
\hat{M}(x)=
-\gamma i \int dz \sum_{kl}\left(
w(z,-q_{kl})\hat{L}_t(z)_{kl}E_{kk}-
w(z-x+\Delta ,-q_{kl})\hat{L}_t(z)_{kl}E_{kl}\right).
\label{fin}
\end{equation} 
Note that this expression literally coincides with (\ref{Mst}) if we change
$\hat{L}_t$ for $L$. Since at $t=0$ the operators $\hat{L}$ and $L$ 
are equal to $L_0$, they coincide for any $t$.

\section{Conclusion}
We have proved that the elliptic Ruijsenaars-Schneider model can be 
obtained by means of the reduction procedure. It is worthwhile to point 
out that we have used not the Hamiltonian but the Poissonian
reduction technique. Our construction is specified by the choice of 
the trigonometric $r$-matrix for the Poisson structure on HD 
and by fixing the special value of the momentum map.
By varying the r.h.s. of the momentum map equation one can derive 
some other systems.  For instance, it is not difficult to specify the 
momentum map equation in a way that leads to the elliptic Calogero-Moser 
model. It clarifies the coincidence of the dynamical $r$-matrices for 
these two models pointed out in \cite{Sur1}.

We have considered the simplest example of HD for $\widetilde{GL(N)}$.
It seems to be interesting to examine the Poissonian reductions  of HD that 
correspond to some other choices of Lie groups or $r$-matrices.

{\bf ACKNOWLEDGMENT} 
The authors are grateful to L.Chekhov,  A.Gorsky  and  N.A.Slavnov
for valuable discussions. This work is supported in part by the RFFR
grants N96-01-00608 and N96-01-00551 and by the ISF grant a96-1516.


\begin{thebibliography}{99}
{\small
\bibitem{Skl} Sklyanin E.K.: Alg.Anal.,  6(2) (1994) 227.
\bibitem{ABT} Avan J., Babelon O. and Talon M.: Alg.Anal. 6(2) (1994) 67.
\bibitem{BS} Braden H.W. and Suzuki T.: Lett.Math.Phys. 30 (1994) 147.
\bibitem{BAB} Billey E., Avan J. and Babelon O.: Phys.Lett. A 188 (1994) 263.  
\bibitem{AR} Avan J. and Rollet G.: The classical r-matrix for the 
relativistic Ruijsenaars-Schneider system, preprint BROWN-HET-1014 (1995).  
\bibitem{AM} Arutyunov G.E. and Medvedev P.B.: 
Generating equation for $r$-matrices related to dynamical
systems of Calogero type; Preprint hep-th/9511070, to appear in Phys.Lett.A.
\bibitem{Sur1} Yu.B.Suris, Why are the rational and hyperbolic
Ruijsenaars-Schneider hierarchies governed by the same $R$-matrix
as the Calogero-Moser ones ? hep-th/9602160;
Elliptic Ruijsenaars-Schneider and 
Calogero-Moser hierarchies are governed by the same $r$-matrix,
solv-int/9603011.
\bibitem{NKSR} Nijhoff F.W., Kuznetsov V.B., Sklyanin E.K. and Ragnisco O.:
Dynamical $r$-matrix for the elliptic Ruijsenaars-Schneider model;
Preprint, March, solv-int/9603006.
\bibitem{AT} Avan J. and Talon M., Phys.Lett.B303 (1993) 33-37.
\bibitem{Arn} V.I.Arnol'd, Mathematical methods of classical mechanics,
Graduate Texts in Math., 60, Spinger-Verlag, Berlin, New-York, 1989.
\bibitem{OP} M.A.Olshanetsky, A.M.Perelomov, Invent. Math.  37 (1976) 93.
\bibitem{KKS} Kazhdan D., Kostant B., Sternberg S.,
Comm.Pure Appl.Math. 31 (1978) 481.
\bibitem{OP1} M.A.Olshanetsky, A.M.Perelomov, Phys. Reps. 71 (1981) 313.
\bibitem{OP2} M.A.Olshanetsky, A.M.Perelomov, Phys. Reps. 94 (1983) 6.
\bibitem{R} Ruijsenaars S.N.: Comm.Math.Phys. 110 (1987) 191.
\bibitem{Sem} Semenov-Tian-Shansky M.A.: Teor.Math.Phys. 93 (1992) 302 (in 
Russian).
\bibitem{GN} Gorsky A. and Nekrasov N.,  Nucl.Phys. B414 (1994) 213;
Nucl.Phys. B436 (1995) 582; A.Gorsky, 
Integrable many body systems in the field theories, 
Prep. UUITP-16/94, (1994).
\bibitem{GNH} Gorsky A. and Nekrasov N.: Elliptic Calogero-Moser
system from two-dimensional current algebra; Preprint hep-th/9401021.
\bibitem{SR} Reshetikhin N.Yu. and Semenov-Tian-Shansky M.A.: 
Lett.Math.Phys. 19 (1990) 133-142.
}
\end{thebibliography}
\end{document}